\documentstyle[aps,prd,eqsecnum,twocolumn,epsf]{revtex}

\newcommand{\beq}{\begin{equation}}
\newcommand{\beqn}{\begin{eqnarray}} 
\newcommand{\eeq}{\end{equation}}
\newcommand{\eeqn}{\end{eqnarray}}

\def\beq{\begin{equation}}
\def\eeq{\end{equation}}

\newcommand{\lsim}{\mbox{\raisebox{-1.ex}{$\stackrel
     {\textstyle<}{\textstyle \sim}$}}}
\newcommand{\square}{\kern1pt\vbox{\hrule height
1.2pt\hbox{\vrule width 1.2pt\hskip 3pt
   \vbox{\vskip 6pt}\hskip 3pt\vrule width 0.6pt}\hrule
height 0.6pt}\kern1pt}

\begin{document}
\draft \twocolumn[\hsize\textwidth\columnwidth\hsize\csname
@twocolumnfalse\endcsname

\title{Inflationary preheating and primordial black holes}
\author{Bruce A. Bassett$^*$  and Shinji Tsujikawa$^{\P}$}
\address{$^*$ Relativity and Cosmology Group (RCG), University of
Portsmouth, Mercantile House, Portsmouth,  PO1 2EG, 
England\\[.3em]}
\address{$^{\P}$ Department of Physics, Waseda University,
3-4-1 Ohkubo, Shinjuku-ku, Tokyo 169-8555, Japan\\[.3em]}
\date{\today}
\maketitle
\begin{abstract}
Preheating after inflation may over-produce primordial black holes (PBH's)
in many regions of parameter space. 
As an example we study two-field models with a 
massless self-interacting inflaton, taking into account second order 
field and metric backreaction effects as spatial averages.
We find that a complex quilt of 
parameter regions above the Gaussian PBH over-production 
threshold emerges  due to the enhancement of 
curvature perturbations on all scales. 
It should be possible to constrain realistic models of inflation
through PBH over-production although many issues,
such as rescattering and non-Gaussianity, remain 
unsolved or unexplored.
\end{abstract}
\vskip 1pc 
\centerline{RCG-00/27,  WUAP-00/22}
\vskip 2pc
]


\underline{\em Introduction} -- The issue of whether initial 
conditions at the Planck
era were suitable for the onset of inflation is both complex and 
controversial  \cite{init,GPVT}. With these subtleties aside, there
remains a cavernous space of possible inflationary models \cite{LR}. 
The requirement of graceful exit from the cold inflationary phase 
into an acceptable  radiation-dominated  FLRW universe has proven 
a powerful filter on this model space.

Failure to exit gracefully spelt the end of the 
old inflationary scenario \cite{GW}, is perhaps the major stumbling block 
in pre-big-bang  models\cite{PBB} and continues to plague string and 
supergravity models of inflation through the threat of overproduction of 
dangerous relics such as moduli and gravitinos \cite{relic}.
 
Perhaps the most radical way to end inflation is via preheating
(see e.g., \cite{pre}) in which runaway particle production
occurs in fields coupled non-gravitationally to the inflaton. 
This explosive growth of quantum fluctuations drives similar resonances in 
metric perturbations on scales which range from  cosmological to  
sub-Hubble \cite{mpre1}. 

It is now recognized that in certain models preheating can alter
the predictions of inflation for 
the Cosmic Microwave Background
(CMB)\cite{mpre2,mpre3,mpre4,mpre5,mpre6} by
exponentially amplifying  super-Hubble metric perturbations. 
This does not violate causality  but  depends sensitively on the preceding 
inflationary phase which determines the spectrum of $\chi$ fluctuations 
\cite{JS,Ivanov,LLKW,HM,shinji}.
In this paper we discuss what appears to be a more 
robust mechanism for constraining models of preheating -- 
over-production of primordial black holes (PBH's).

The idea that the amplification
of metric perturbations during preheating  would lead to enhancement
of PBH abundances was raised early on \cite{mpre1} and has been 
alluded to frequently since; e.g., \cite{JS,PE}. 
Recently Green and Malik 
\cite{GM} have used a semi-analytic approach which incorporates second 
order $\chi$ field fluctuations  to study PBH formation in a 
two-field massive inflaton model. 

Their results suggest that during strong preheating ($q \gg 1$ 
\cite{pre}), PBH formation could violate astrophysical limits 
before backreaction  ends the resonant growth of $\chi$ fluctuations. 
This is a crucial issue since strong preheating is generic in many models 
of inflation.  However, Green and Malik used the results of \cite{pre} for the 
estimate of the time at which 
backreaction ends the initial resonance. As they point out this 
estimate does not include  metric perturbations or rescattering and hence 
could be misleading. 

\begin{figure}
\epsfxsize=3.2in
\epsffile{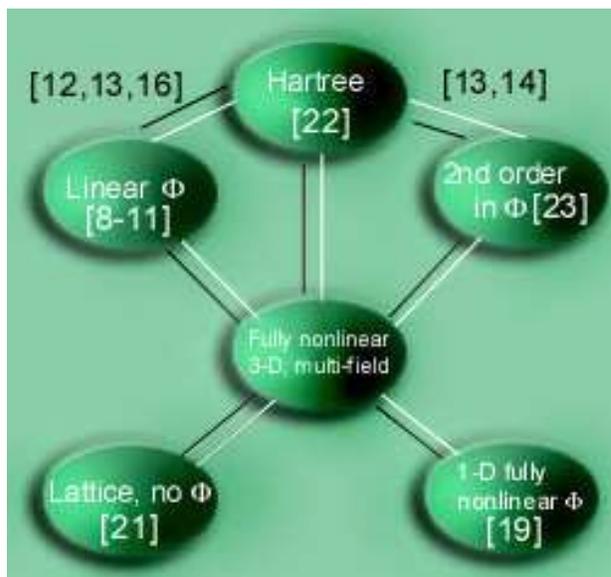}
\caption{A schematic figure showing the numerical approaches to preheating 
with numbers in brackets denoting appropriate  references. See the text for
discussion.}
\end{figure}

Here we present first estimates of PBH production including backreaction
computed dynamically. We find that while preheating may lead to over-production
of PBH's in some regions of parameter space, the result is sensitive to 
many subtle issues. 

To place our methods in context, consider 
Fig.~1 which  shows the different  numerical studies of 
preheating undertaken in the literature. The eventual  goal 
of these studies is a fully
nonlinear analysis of multi-field preheating including metric perturbations. 
So far this has been achieved without metric
perturbations (``no $\Phi$'') - often with simplified expansion dynamics -
through lattice simulations \cite{lattice}.  The furthest the community has
progressed \cite{PE} in solving the full Einstein field 
equations is in a model with plane wave symmetry  and a  
single scalar field. 

An alternative to full lattice simulations of preheating is the use of the 
Hartree, large-N, and mean field approximations \cite{hartree}. 
Recently the Hartree approximation  has been combined with the linear  
approximation for metric perturbations $\Phi$ \cite{mpre5,HM,shinji}
and, in \cite{mpre6,JS}, with the second order metric 
perturbations formalism of 
Abramo {\em et al.} \cite{ABM}. It is this latter approach that we adopt.

Immediate goals  are fully
nonlinear spherically symmetric simulations suitable for studying individual
PBH formation (c.f. \cite{boson}) and inclusion of rescattering effects
in the presence of metric perturbations, $\Phi$. The latter requires 
going beyond the Hartree approximation and evaluating double and 
triple convolutions.  

\underline{\em The Model} -- We consider the two scalar field 
chaotic inflation model 
\begin{eqnarray}
V(\phi, \chi) = \frac{\lambda}{4}\phi^4 + \frac{g^2}{2}\phi^2 \chi^2,
\label{potential}
\end{eqnarray}
where $\phi$ is an inflaton field.
During inflation $\chi$ decreases rapidly towards zero if $g^2/\lambda 
\gg 1$ in which case the temperature anisotropies 
in the CMB simply scale as 
$\Delta T/T \sim \sqrt{\lambda}$. We therefore choose a self-coupling of 
$\lambda = 10^{-13}$. During preheating, $\chi$ and $\delta\chi_k$ grow 
exponentially in very specific geometric channels or resonance bands which
are well understood in terms of Floquet theory\cite{GKLS97,mpre3}. 

We assume a flat background FLRW geometry with perturbations in 
the longitudinal gauge\cite{mpre1}:
\begin{eqnarray}
ds^2=-(1+2\Phi)dt^2
+a^2(1-2\Phi)\delta_{ij} dx^i dx^j,
\label{B2}
\end{eqnarray}
where $\Phi = \Phi({\bf x},t)$, the natural generalization of the 
Newtonian potential, describes scalar perturbations  and  
$a = a(t)$ is the scale factor.
We decompose the scalar fields into homogeneous parts and fluctuations
as $\phi(t, {\bf x}) \to \phi (t)+\delta\phi (t,{\bf x})$ and 
$\chi(t, {\bf x}) \to \chi (t)+\delta\chi (t,{\bf x})$.

The structure of the linearized Einstein field equations for this system 
can be schematically written in terms of two vectors: one for the FLRW 
background dynamics ${\bf X} = (\phi, \dot{\phi}, \chi, \dot{\chi}, 
a, \dot{a})$, and one for the perturbation variables in Fourier 
space: ${\bf Y}_k = (\delta\phi_k, \delta\dot{\phi}_k,
\delta \chi_k, \delta\dot{\chi}_k, \Phi_k, \zeta_k)$.

While we solve the system of linearized Einstein field equations in the 
longitudinal gauge, it is convenient to calculate PBH constraints 
in terms of the curvature perturbation $\zeta_k$ rather than
$\Phi_k$. 
$\zeta_k$ is defined in terms of $\Phi_k$
and the Hubble parameter, $H \equiv \dot{a}/a$, by
\begin{eqnarray}
\zeta_k \equiv \Phi_k-\frac{H}{\dot{H}}\left(H\Phi_k+\dot{\Phi}_k\right),
\label{C1}
\end{eqnarray}
and is usually conserved on super-Hubble scales 
in the adiabatic single field inflationary scenario.
In the multi-field case which we consider in this paper, 
this quantity can change nonadiabatically due to the amplification
of isocurvature (entropy) perturbations.

\begin{figure}
\epsfxsize=3.5in
\epsffile{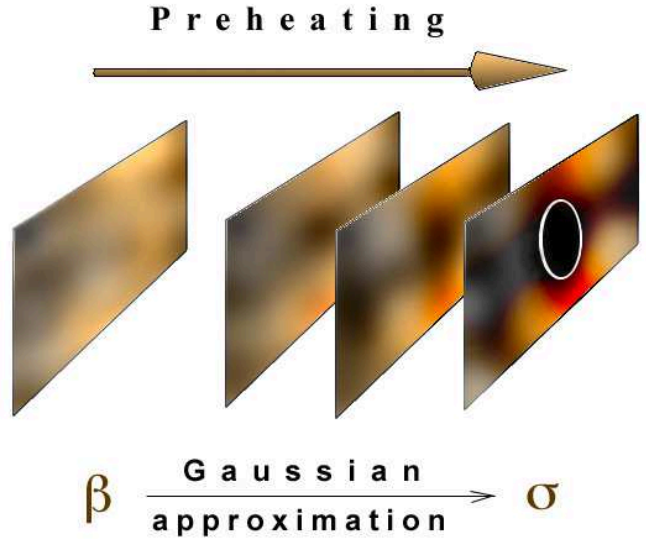}
\caption{An
illustration of primordial black hole (PBH) formation during preheating due
to growth of density perturbations.  The PBH event horizon 
is schematically shown  by the white ring in the final panel. 
Astrophysical limits on PBH's constrain  $\beta$, 
the ratio of PBH to total energy density. 
To constrain theory one needs to map $\beta$ into the 
mass variance $\sigma$, most easily achieved with a Gaussian 
or chi-squared assumption for density perturbations. It 
is $\sigma$ that we calculate in our  simulations.}
\end{figure}

We include backreaction effects to second order in {\em both} field and 
metric perturbations\cite{ABM},
which implies that we integrate coupled integro-differential equations.  
The precise structure of these equations and additional details 
can be found in Appendix and \cite{mpre5,mpre6,JS,ABM}. Here we 
illustrate the skeletal structure of the system, which has the form
\beqn
\dot{{\bf X}} &=& {\bf F}({\bf X},\langle{\bf Y}^2\rangle), \nonumber \\
\dot{{\bf Y}}_k &=& {\bf G}({\bf X},\langle{\bf Y}^2\rangle)~{\bf Y}_k, 
\nonumber \\
{\bf F} &=& {\bf F}_{\rm hom} + {\bf F}_{\rm pert}, \nonumber \\
{\bf F}_{\rm pert} &=& {\bf F}_{\rm pert}(\langle \delta\phi^2 \rangle, 
\langle \delta\chi^2 \rangle, \langle \Phi^2 \rangle, ...)\,, 
\label{sys}
\eeqn
where the variance is defined by
\beq
\langle \diamond^2 \rangle \equiv \frac{1}{2\pi^2}
\int k^2 |\diamond|_k^2 dk \,, 
\eeq
for any field $\diamond$. ${\bf F}$ and ${\bf G}$ are nonlinear functions
of the spatially homogeneous background
vector ${\bf X}$ and the variances of the components of ${\bf Y}$.
The complete system is integrated from 50 e-folds before the end of inflation 
to provide the  appropriate initial conditions for preheating. 
The initial values at the start of inflation  are chosen as
$\phi=4m_{\rm pl}$ and $\chi=10^{-3}m_{\rm pl}$
with conformal vacuum states for the 
fluctuations\footnote{Using the initial condition $\chi=10^{-6}m_{\rm pl}$
we reproduced the results of Ref.~\cite{mpre6}.}.
Including the field variances ensures total energy  conservation at 1-loop.

\underline{\em Primordial black hole constraints} -- Since PBH's form 
from large density fluctuations\cite{carr},
it is an obvious concern that preheating might encounter 
problems with PBH constraints arising from the Hawking 
evaporation of small PBH's or from overclosure of the universe
($\Omega_{\rm PBH} > 1$) for heavy PBH's. 

To quantify this suspicion one needs to compute the mass function 
$\beta$\cite{BP,gl}: 
\begin{eqnarray}
\beta = \frac{\rho_{\rm PBH}}{\rho_{\rm TOT}}
=\int_{\delta_c}^{\infty} P(\delta)~d\delta,
\label{D1}
\end{eqnarray}
where $P(\delta)$ is the probability distribution of the density 
contrast, $\delta$, and $\delta_c$, $(\approx 0.7)$ \cite{nj}, 
is the critical value at which 
PBH formation occurs in the radiation dominated era.

Usually one assumes a Gaussian distribution 
$P(\delta)=1/(\sqrt{2\pi}\sigma) \exp[-\delta^2/(2\sigma^2)]$,
where $\sigma$ is the mass variance at horizon crossing.
Observational constraints imply that
$\beta < 10^{-20}$ over a very wide range of mass scales, which translates
into a bound on the mass variance of $\sigma<\sigma_{*}=0.08$. $\sigma 
> \sigma_*$ corresponds to PBH over-production in the Gaussian 
distributed case. 
When the distribution is instead first order chi-squared -- an approximation 
to the $\chi$ density fluctuations in  
preheating  (see the later discussions) -- the threshold is 
$\sigma_{*}=0.03$\cite{GM}.

Defining the power-spectrum of the curvature perturbation as 
${\cal P}_{\zeta} \equiv k^3|\zeta_k|^2/(2\pi^2)$,  
the mass variance can be expressed as\cite{GM,LL}
\begin{eqnarray}
\sigma^2=\left(\frac49\right)^2 \int_0^{\infty}
\left(\frac{k}{aH}\right)^4 {\cal P}_{\zeta}
\tilde{W}(kR) \frac{dk}{k}.
\label{D3}
\end{eqnarray}
We choose a Gaussian-filtered  window function 
$\tilde{W}(kR) \equiv \exp(-k^2 R^2/2)$ where $R \equiv 1/k_*$ is 
the artificial smoothing scale \cite{LL}.
We can expect exponential increase of $\sigma$ due to the excitement 
of field and metric perturbations during preheating.
We solved the Einstein equations (\ref{sys}) numerically, varying the 
ratio $g^2/\lambda$, and evaluated the mass variance with two cut-offs 
$k_*=aH$ and $k_*=10aH$ to investigate sensitivity to cut-off effects.

\begin{figure}
\epsfxsize=3.5in
\epsffile{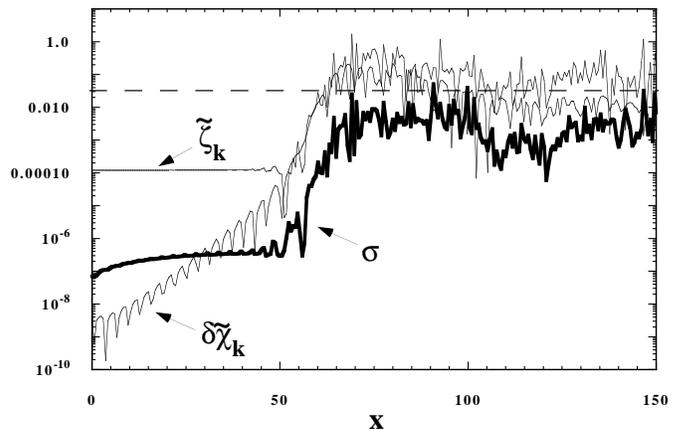}
\caption{Threshold PBH formation -- the growth of $\sigma$, 
$\tilde{\zeta}_k \equiv k^{3/2}\zeta_k$, and 
$\delta\tilde{\chi}_k \equiv k^{3/2}\delta\chi_k/m_{\rm pl}$ 
for a super-Hubble mode $\kappa \equiv k/(\sqrt{\lambda}\phi_0)
=10^{-22}$ vs dimensionless time $x \equiv \sqrt{\lambda}\phi_0 \eta$ 
in the case $g^2/\lambda = 2.5$, where $\phi_0 
\simeq 0.1 m_{\rm pl}$ is the value of inflaton when it begins to oscillate coherently.
With the choice $k_* = aH$ in the window function 
$\tilde{W}(kR)=\exp(-k^2R^2/2)$, $\sigma$ just
reaches the threshold $\sigma_{*}=0.03$ for the PBH formation
for chi-squared first order distributions.}
\label{sigmaevolution}
\end{figure}

When $\chi_k$ fluctuations are amplified during preheating,
this stimulates the growth of the metric perturbation, $\Phi_k$.
On cosmological scales this effect is sensitive to the 
suppression of $\chi$  and $\delta\chi_k$ modes in the preceding inflationary
phase. 

When $g^2/\lambda={\cal O}(1)$, this suppression is weak since the 
$\chi$ field is light \cite{mpre3}
and once the long-wave $\delta\chi_k$ modes grow to of order
$\delta\phi_k$ during preheating, super-Hubble $\Phi_k$ and 
$\zeta_k$ are  amplified until backreaction effects
shut off the resonance. This amplification occurs 
in the region  
$1<g^2/\lambda<3$\cite{mpre3,mpre4,mpre5,mpre6}, 
where the $k \simeq 0$ modes lie in a resonance band. 
The increase in  $\zeta_k$ leads to a corresponding growth of 
the mass variance
$\sigma$ which can reach the threshold 
$\sigma_{*}=0.03$ for $1<g^2/\lambda<3$ and $6<g^2/\lambda<10$
with the cut-off set at $k_*=aH$, i.e., around the Hubble scale 
(see Fig.~\ref{sigmaevolution}).

\begin{figure}
\epsfxsize=3.5in
\epsffile{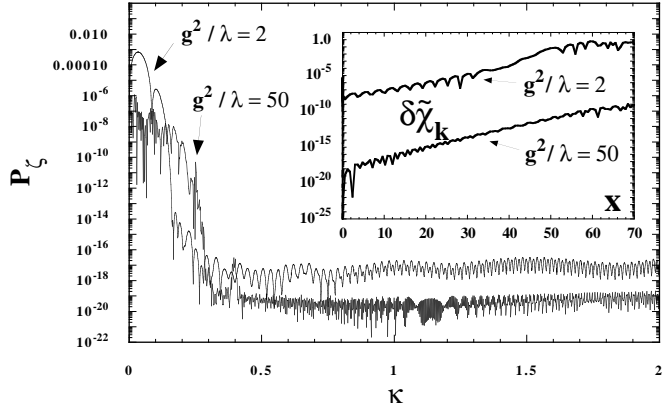}
\caption{The power spectrum of $\zeta$ at the end of preheating 
for the values of $g^2/\lambda = 2, 50$ with $k_*=aH$. 
Since the inflationary suppression of $\chi$ becomes stronger as
$g^2/\lambda$ is increased, the growth of $\zeta_k$ at long wavelengths 
is suppressed. Note also the dominance of the $g^2/\lambda = 2$ 
modes at sub-Hubble scales $\kappa > 1$. 
{\bf Inset :} The evolution of $\delta\chi_k$ for a super-Hubble mode 
$\kappa \equiv k/(\sqrt{\lambda}\phi_0)=10^{-22}$ for $g^2/\lambda = 2, 50$. 
The suppression of the initial conditions, due to the preceding inflationary 
phase, in the heavy case  $g^2/\lambda = 50$ is evident.}
\label{spectrum}
\end{figure}

As $g^2/\lambda$ is increased, the $\chi$ field becomes heavy and 
suppressed during inflation. 
This restricts the amplification of super-Hubble metric perturbations 
\cite{mpre6} despite the fact that  
the $k \rightarrow 0$ mode of $\delta\chi_k$ lies 
in a resonance band for 
$n(2n-1)<g^2/\lambda<n(2n+1)$, $n = 1,2,3...$ 
\cite{GKLS97}, as is evident from  Fig.~\ref{spectrum}.
However, since sub-Hubble $\delta\chi_k$ modes are not  
suppressed during inflation\cite{JS,mpre2}, $\Phi_k$ and 
$\zeta_k$ on sub-Hubble
scales do exhibit nonadiabatic, resonant, growth for $g^2/\lambda \gg 1$
\footnote{We have reproduced the result that the homogeneous part 
of the $\chi$ field is amplified  by the second order couplings between 
$\Phi_k$ and $\delta\chi_k$\cite{JS} despite of the 
inflationary suppression.}, 
which leads to growth of $\sigma$.

\begin{figure}
\epsfxsize=3.5in
\epsffile{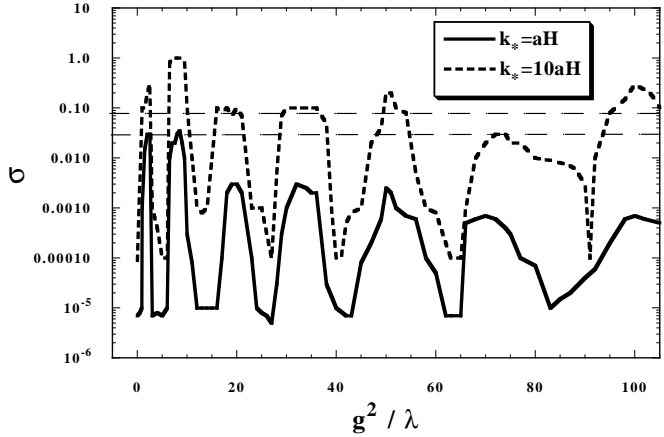}
\caption{The mass variance $\sigma$ vs $g^2/\lambda$ 
for two window function cut-offs $k_* = aH$ and $k_* = 10aH$. 
The threshold of $\sigma_{*} = 0.03$ in the chi-squared
distributed case is shown and is marginally crossed for the regions around 
$g^2/\lambda \sim 2$ and $g^2/\lambda \sim 8$ when $k_* = aH$. 
For $k_* = 10aH$ a quilt of regions above the Gaussian threshold 
$\sigma_{*} = 0.08$ emerges 
which coincide closely with $g^2/\lambda = 2n^2$ corresponding to Floquet 
indices with maxima at longest wavelengths. }
\label{sigma}
\end{figure}

However, we do {\em not} find that this is significant enough to lead to 
$\sigma > \sigma_*$ for $g^2/\lambda \gg 1$, except for very short intervals
around $\dot{\phi} = 0$, in contrast to the 
expectations of \cite{GM}. However, when we enlarge the cut-off 
frequency $k_*$ to $10aH$,  we do find  $\sigma > 0.08$ in wide
ranges of parameter space (see Fig.~\ref{sigma}).
Somewhat surprisingly, these super-threshold regions are 
all clustered  around the  super-Hubble resonance bands in $g^2/\lambda$ 
space.

\underline{Initial conditions for the $\chi$ field} -- An issue 
of general importance which has been little studied is that of initial
conditions for non-inflaton fields at the {\em start} of inflation. 
In our model these fields are represented by $\chi$ and the initial value 
is set $55$ e-folds before the end of inflation. 
This problem has two facets - the initial value of the background, or vacuum 
expectation value of $\chi$, and the initial value of the distribution of 
fluctuations, i.e., $\delta\chi_k$. 

A sensible choice for the latter is the Bunch-Davis vacuum, but it is the 
initial value of the homogeneous part of $\chi$, which is of the most 
importance, since if 
$\chi = 0$ (the minimum of the potential) no resonance can
occur at linear order. 

We have found four suggestions for setting the initial value, $\chi_i$,
as follows.

(1)  Choose the value of $\chi$ which maximises the 
probability distribution in eternal
inflation for fixed large values of the inflaton ($\phi > 1 m_{\rm pl}$) at 
a specific time. 
Since the regions with the largest Hubble constant dominate the distribution 
\cite{eternal} this corresponds to choosing $\chi > 1 m_{\rm pl}$, i.e.,  
super-Planckian chaotic initial conditions for $\chi$.

This suggestion is, however, sensitive to the choice of a hypersurface 
for setting the initial  conditions. If one defines initial 
conditions on the hypersurface of energy
density equal to the Planck energy for instance, then the Hubble constant
will likely be maximised by placing all energy into the field with the flatest
potential, rather than distributing it amoung the various fields, 
some of which may  have steep potentials. This will lead to vanishingly small 
initial $\chi$ unless $\chi$ is a good inflaton, i.e., $g^2~\lsim~\lambda$.

(2) Choose $\chi_i$ to  satisfy $\chi_i^2 = \langle \chi^2 
\rangle$ \cite{mpre6}. If we use the Bunch-Davis condition, $\chi_k 
\sim 1/\sqrt{\omega_k} \sim 1/\sqrt{g\phi}$, we can estimate 
$\langle \chi^2 \rangle = \frac{1}{2\pi^2}\int k^2 |\chi_k|^2 dk$ to be 
$\sim k_*^3/(g\phi)$ 
if the variance is super-Hubble dominated during inflation and where $k_* = 
H$ is the natural cut-off at the Hubble scale.

Now if inflation is driven by $\phi$ then $H^2 \simeq V(\phi)/m_{\rm pl}^2$ and
we find $\chi_i^2 \simeq V(\phi)^{3/2}/(g\phi m_{\rm pl}^3)$. For the 
potential (\ref{potential}) this 
leads to the estimates $\chi_i \sim 10^{-5} m_{\rm pl}$ for $g^2/\lambda = 
{\cal O}(1)$ and $\chi_i \sim 10^{-8} g^{-1/2} m_{\rm pl}$ 
for $g^2/\lambda \gg 1$ 
if we take $\lambda \sim 10^{-13}$ and $\phi_i \sim 4m_{\rm pl}$.

(3) Choose the value of $\chi_i$ which leads to a stationary distribution 
in eternal inflation (where the classical drift and quantum fluctuations 
are balanced) \footnote{We thank Alan Guth for this suggestion.}. 
Assuming quantum fluctuations $\delta\phi \sim H/2\pi$ on 
characteristic time scales $\delta t \sim H^{-1}$ one arrives at 
$\chi_i \sim H^3/(g^2\phi^2) \sim \lambda^{3/2} \phi^4/(g^2m_{\rm pl}^3)$ 
and hence $\chi_i \sim 10^{-4}-10^{-3} m_{\rm pl}$ 
for $g^2/\lambda = {\cal O}(1)$ and 
$\chi_i \sim 10^{-16} g^{-2} m_{\rm pl}$ for $g^2/\lambda \gg 1$.

(4) Finally we may choose the value of $\chi_i$ which corresponds to the 
instantaneous minimum of the   potential. It suggests $\chi_i = 0$. 
This argument has several problems the most fundamental of which is that 
the system is not in equilibrium since the $\chi$ field is not strongly 
coupled except for $g^2/\lambda \gg 1$. 

\begin{figure}
\epsfxsize=3.5in 
\epsffile{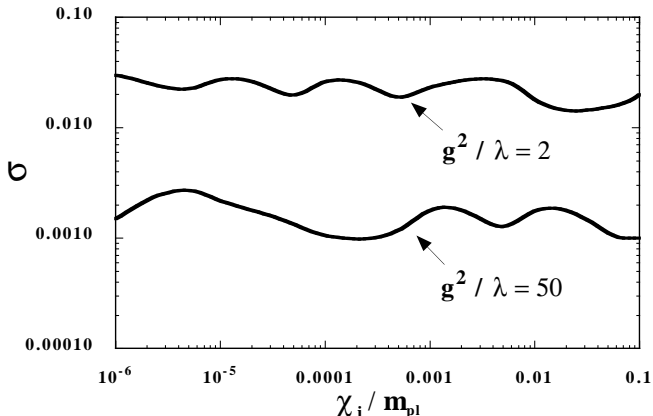}
\caption{The smoothed dependence of the final mass variance $\sigma$ on 
the initial condition, $\chi_i$, 55 e-folds before the start of preheating 
for  $g^2/\lambda = 2, 50$. Note the relatively weak dependence 
in both cases. }
\label{initial}
\end{figure}

Despite the wide range of possible initial values, $\chi_i$, at the start of 
inflation, Fig.~\ref{initial} shows that the final mass variance, and 
hence the probability of PBH over-production, depends rather weakly on 
$\chi_i$.

\underline{\em Potential problems and unresolved issues} -- Our results
suggest that PBH over-production  may {\em not} be generic in 
strong preheating. 
However they can only be considered as preliminary for a number of reasons:

\begin{itemize}
\item There are at least two fields critically involved in preheating. 
Even if the inflationary fluctuations are Gaussian, the fluctuations 
induced by preheating are typically not. If the $\chi$ field has no 
vacuum expectation value, its  density fluctuations are roughly 
$\propto \delta\chi^2$, so approximately chi-squared if $\delta\chi$ is 
Gaussian distributed. As discussed above, we take $\chi \neq 0$.  The 
recent results of \cite{latticeeasy} suggest that $\chi$ is Gaussian 
distributed before rescattering sets in  
and hence the density perturbations would be 
Gaussian, at least while dominated by linear fluctuations. 

Rescattering leads to non-Gaussian distributions and to  
$\delta \phi \propto \delta\chi^2$ \cite{pre}. The applicability of the 
criterion $\sigma > \sigma_*$ therefore depends largely on when PBH formation
actually takes place - before or after rescattering. 

Further, the density fluctuations may go nonlinear.  Since 
$\delta \in [-1,\infty)$ this necessarily 
skews the distribution, similar to the toy-model discussed in the second 
Ref. of \cite{BP}. Non-Gaussianity may drastically alter the relationship 
between $\beta$ and $\sigma$ \cite{BP,iv}, changing $\sigma_*$ and 
requiring the use of higher-order statistics. 

\item In preheating the Hubble radius is vastly smaller than the true particle 
horizon and  resonance bands often cover the complete range of scales. 
Predicting the mass spectrum of 
PBH's created during preheating is therefore a subtle issue. 
Crudely one expects a wide  range of PBH masses to be produced, 
even without criticality  arguments \cite{nj}. 

This is related to our results showing   
cut-off, $k_*$, sensitivity. The increase in $\sigma$ when  $k_*$ is
altered from $aH$ to $10aH$ reflects the important contributions of 
sub-Hubble modes.  Does this necessarily imply that the  resulting 
PBH's are very small ? If so, they are not constrained 
since they  evaporate harmlessly long before nucleosynthesis.


\item We have not included rescattering. This is known to enhance
variances  over the Hartree approximation at small
resonance parameters, $q$,  in the absence 
of metric perturbations\cite{lattice}. 
For $q \gg 1$ however, the situation is reversed 
and variances are overestimated by the Hartree approximation. Whether these results are stable to inclusion of metric perturbations is unknown, but this 
may provide a way to avoid PBH over-production since it should
filter through to $\zeta_k$ and $\sigma$.

%

\item Fig.~\ref{sigma} shows $\sigma$ as a function 
of $g^2/\lambda$. The value of $\sigma$ plotted is its maximum at the 
end of preheating. However, $\sigma$ does  grow larger than this value, 
instantaneously  exceeding $0.01$, even for $k_* = aH$, when 
$\dot{\phi} = 0$. We choose the 
more conservative route of not taking these as the true maxima, but the 
question remains, can large $\sigma$, attained for very short periods, lead to
PBH formation ?  

\item We solved the $\chi$ field equation, including second
order terms such as $\langle\Phi\delta\ddot{\chi}\rangle$\cite{JS}.
Initially $\delta\chi$ and $\Phi$ are correlated but when the $\chi$ 
fluctuations are sufficiently amplified, they are well 
described by classical stochastic waves\cite{HM}, which may be uncorrelated
with metric variables. 

It is uncertain that contributions of second order
metric terms should be included during such classical regimes. 
Since this issue affects $\chi$ rather significantly, 
the quantum to classical transition appears to be of quantitative importance, 
deserving further study.

\end{itemize}

\underline{\em Conclusions} -- We have studied primordial black 
hole (PBH) formation during 
preheating using numerical simulations of the perturbed Einstein field
equations including second order field and metric backreaction effects. 
We found that there exist parameter ranges where standard
Gaussian and chi-squared thresholds for PBH formation are exceeded.

Nevertheless,  the results are not unambiguous. We discovered a 
significant sensitivity to the window function cut-off, $k_*$, and since
preheating is expected to lead to non-Gaussian fluctuations, it is not 
clear how realistic the Gaussian threshold for PBH formation is. 
Nevertheless, PBH over-production constraints are very robust. The 
study of PBH's in preheating is an exciting area which may lead 
to strong constraints on realistic inflationary models. 

We note that there are a number of possible escape routes 
to preserve preheating but avoid PBH over-production.
Fermionic preheating is very unlikely to lead to PBH formation unless 
the fermions are extremely massive. Similarly, instant preheating 
\cite{instant}, which
draws energy away from the $\chi$ field almost immediately seems likely 
to stall PBH formation, as does a large $\chi$ self-interaction. 

On the other hand, since growth of $\zeta_k$ and $\sigma$ is seeded 
through isocurvature/entropy 
perturbations \cite{LLKW}, it is possible that other models of reheating,
such as non-oscillatory models \cite{no}, which lead to significant 
isocurvature modes, may also have a PBH over-production problem. 

Nevertheless, the precise scenario of the PBH formation during preheating 
can only be understood properly by overcoming two serious hurdles - (1) 
understanding the probability distribution of density fluctuations 
during preheating and (ii) going to fully
nonlinear simulations of resonant PBH formation which include 
rescattering and nonlinear metric perturbations.

\underline{\em Acknowledgements} -- We thank Christopher Gordon, 
Alan Guth, Anne Green, Roy Maartens, 
Kei-ichi Maeda, Karim Malik,  Masaaki Morita, and Takashi Torii 
for useful discussions and comments.

\section*{Appendix: Detailed form of the evolution equations}

In this Appendix we present the evolution equations in details.
We include second order field and metric backreaction 
effects\cite{ABM} in the background equations,
which is combined with the Hartree approximation\cite{hartree}.

Then the Hubble parameter 
and homogeneous parts of the scalar fields satisfy\cite{mpre6,JS}:
\begin{eqnarray}
 H^2 &=&
  \frac{8\pi G}{3}
     \Biggl[ \frac12 \dot{\phi}^2+ \frac12
    \langle \delta \dot{\phi}^2 \rangle+
     \frac{1}{2a^2} \langle (\nabla \delta\phi)^2 
      \rangle  \nonumber \\
     &+& \frac12 \dot{\chi}^2+ \frac12
     \langle \delta \dot{\chi}^2 \rangle+
     \frac{1}{2a^2} \langle (\nabla \delta\chi)^2 
     \rangle  \nonumber \\
      &+& \frac14 \lambda(\phi^4+6\phi^2
      \langle \delta\phi^2 \rangle
      +3\langle \delta\phi^2 \rangle^2) 
      +\frac12 g^2\phi^2\langle\chi^2\rangle
       \nonumber \\
      &+& 2(\lambda\phi^3+g^2\phi\chi^2) 
      \langle \Phi\delta\phi \rangle+
      2g^2\phi^2\chi \langle \Phi\delta\chi \rangle \Biggr]
      \nonumber \\
      &+& 4H\langle\Phi\dot{\Phi}\rangle-
      \langle \dot{\Phi}^2 \rangle+\frac{3}{a^2}
       \langle(\nabla \Phi)^2\rangle,
\label{B6}
\end{eqnarray}
\begin{eqnarray}
(\ddot{\phi} &+& 3H \dot{\phi})(1+4\langle\Phi^2\rangle)
+\lambda \phi(\phi^2+3 \langle \delta\phi^2 \rangle)  
\nonumber \\
&+& g^2(\chi^2+\langle \delta\chi^2 \rangle)\phi
-2\langle\Phi\delta\ddot{\phi}\rangle
-4\langle\dot{\Phi}\delta\dot{\phi}\rangle
 \nonumber \\
&-& 6H\langle\Phi\delta\dot{\phi}\rangle
+4\dot{\phi}\langle\dot{\Phi}\Phi\rangle
-\frac{2}{a^2} \langle \Phi\nabla^2(\delta\phi)\rangle
=0,
\label{B7}
\end{eqnarray}
\begin{eqnarray}
(\ddot{\chi} &+& 3H \dot{\chi})(1+4\langle\Phi^2\rangle)
+g^2(\phi^2+3 \langle\delta\phi^2 \rangle)\chi   \nonumber \\
&-& 2\langle\Phi\delta\ddot{\chi}\rangle
-4\langle\dot{\Phi}\delta\dot{\chi}\rangle
- 6H\langle\Phi\delta\dot{\chi}\rangle \nonumber \\
&+& 4\dot{\chi}\langle\dot{\Phi}\Phi\rangle
-\frac{2}{a^2} \langle \Phi\nabla^2(\delta\chi)\rangle=0,
\label{B8}
\end{eqnarray}
where $G \equiv m_{\rm pl}^{-2}$ is Newton's 
gravitational constant.
Note that $\langle ... \rangle$ implies a spatial average.
In spite of the exponential suppression during inflation, operative 
when the $\chi$ field is heavy ($g^2/\lambda \gg 1$),
the $\chi$ field can be significantly enhanced in the presence of 
the second order metric backreaction terms in Eq.~(\ref{B8}) 
as pointed out in Ref.~\cite{JS}. 

The Fourier transformed, perturbed Einstein equations
are 
\begin{eqnarray}
\delta\ddot{\phi}_k &+& 3H\delta\dot{\phi}_k
\nonumber \\
&+& \left[\frac{k^2}{a^2}+3\lambda(\phi^2+
\langle\delta\phi^2\rangle) 
+g^2(\chi^2+\langle\delta\chi^2\rangle) \right]
\delta\phi_k \nonumber \\
&=& 4\dot{\phi} \dot{\Phi}_k 
+ 2(\ddot{\phi}
+3H\dot{\phi})\Phi_k-2g^2\phi\chi
\delta\chi_k, 
\label{B3}
\end{eqnarray}
\begin{eqnarray}
\delta\ddot{\chi}_k &+& 3H\delta\dot{\chi}_k+
\left[ \frac{k^2}{a^2}+g^2(\phi^2+\langle\delta\phi^2\rangle)
\right] \delta\chi_k \nonumber \\
&=& 4\dot{\chi} \dot{\Phi}_k 
+ 2(\ddot{\chi}+3H\dot{\chi})\Phi_k-2g^2\phi\chi\delta\phi_k,
\label{B4}
\end{eqnarray}
\begin{eqnarray}
\dot{\Phi}_k+H\Phi_k=4\pi G
(\dot{\phi} \delta\phi_k+\dot{\chi} \delta\chi_k).
\label{B5}
\end{eqnarray}
We find from Eq.~(\ref{B5}) that metric perturbations
grow if $\chi$ and $\delta\chi_k$ fluctuations are amplified 
during preheating and the $\chi$-dependent source term 
exceeds the $\phi$-dependent one. 
When field and metric fluctuations are sufficently amplified, 
the coherent oscillations of the inflaton condensate, $\phi$, are 
destroyed.  The entire spectrum of fluctuations  typically moves out 
of the dominant resonance band and the resonance is  shut off.



\begin{thebibliography}{99}
\bibitem{init} 
S. W. Hawking and N. Turok,  Phys. Lett. B {\bf 425}, 25 (1998);
S. W. Hawking and H. S. Reall, Phys. Rev. D {\bf 59}, 023502 (1999); 
A. Vilenkin, {\it ibid}. {\bf 58}, 067301 (1998);
V. Vanchurin, A. Vilenkin, and S. Winitzki, {\it ibid}. {\bf 61} 083507 (2000).
\bibitem{GPVT} 
D. Goldwirth and T. Piran, Phys. Rev. Lett. {\bf 64}, 2852 (1990); 
T. Vachaspati and  M. Trodden, Phys. Rev. D {\bf 61} 023502 (2000).
\bibitem{LR} 
D. H. Lyth and A. Riotto, Phys. Rep. {\bf 314}, 1 (1999).
\bibitem{GW}
A. H. Guth and E. J. Weinberg, Nucl. Phys. B{\bf 212}, 321 (1983).
\bibitem{PBB}
M. Gasperini, J. Maharana, and G. Veneziano,
Nucl. Phys. B {\bf 472}, 349 (1996).
\bibitem{relic} 
G. F. Giudice, A. Riotto, and I. I. Tkachev, JHEP {\bf 9911}, 
036 (1999); D. H. Lyth, Phys. Lett. B {\bf 469}, 69 (1999); 
R. Kallosh, L. Kofman, A. Linde, and A. Van Proeyen, 
Phys. Rev. D {\bf 61}, 103503 (2000).
\bibitem{pre} 
L. Kofman, A. Linde, and A. A. Starobinsky, Phys. Rev. 
D {\bf 56}, 3258 (1997).
\bibitem{mpre1}
B. A. Bassett, D. I. Kaiser, and R. Maartens, 
Phys. Lett.  B {\bf 455}, 84 (1999); 
B. A. Bassett, F. Tamburini, D. I. Kaiser, and
R. Maartens, Nucl. Phys. B {\bf 561}, 188 (1999).
\bibitem{mpre2}
B. A. Bassett, C. Gordon, R. Maartens,
and D. I. Kaiser, Phys. Rev. D {\bf 61}, 061302 (R) (2000).
\bibitem{mpre3}
B. A. Bassett and F. Viniegra, Phys. Rev. D {\bf 62}, 043507 (2000).
\bibitem{mpre4}
F. Finelli and R. Brandenberger, Phys. Rev. D {\bf 62}, 083502 (2000).
\bibitem{mpre5}
S. Tsujikawa, B. A. Bassett, and F. Viniegra, JHEP {\bf 08}, 019 (2000);
S. Tsujikawa and B. A. Bassett, Phys. Rev. D {\bf 62}, 045310 (2000).
\bibitem{mpre6}
Z. P. Zibin, R. H. Brandenberger, and D. Scott, hep-ph/0007219 (2000).
\bibitem{JS}
K. Jedamzik and G. Sigl, Phys. Rev. D {\bf 61}, 023519 (2000).
\bibitem{Ivanov}
P. Ivanov, Phys. Rev. D {\bf 61}, 023505 (2000).
\bibitem{LLKW}
A. R. Liddle {\em et al.}, Phys. Rev. D {\bf 61}, 103509 (2000).
\bibitem{HM}
A. B. Henriques and R. G. Moorhouse, 
Phys. Rev. D {\bf 62}, 063512 (2000).
\bibitem{shinji}
S. Tsujikawa, JHEP {\bf 07}, 024 (2000).
\bibitem{PE}
M.  Parry and R. Easther, Phys. Rev. D {\bf 62}, 103503  (2000).
\bibitem{GM} 
A. M. Green and K. A. Malik, hep-ph/0008113 (2000).
\bibitem{lattice} 
S. Yu. Khlebnikov and  I. I. Tkachev, Phys. Rev. Lett. {\bf 79}, 
1607 (1997); T. Prokopec and T. G. Roos, Phys. Rev. D {\bf 55}, 
3768 (1997); S. Kasuya and M. Kawasaki, {\it ibid}. {\bf 58},
083516 (1998); M. Parry and A. T. Sornborger, 
{\it ibid}. {\bf 60}, 103504 (1999);
A. Rajantie and E. J. Copeland, Phys. Rev. Lett. {\bf 85},
916 (2000).  
\bibitem{hartree}
S. Yu. Khlebnikov, I. I. Tkachev, Phys. Lett. B {\bf 390} 80 (1997);
S. A. Ramsey and B. L. Hu, Phys. Rev. D {\bf 56}, 678 (1997);
D. Boyanovsky {\em et al.}, {\it ibid}. {\bf 56},  1939 (1997);
S. Tsujikawa, K. Maeda, and T. Torii, {\it ibid}.
{\bf 60}, 063515 (1999); 123505 (1999); 
J. Baacke and C. Patzold, {\it ibid}. {\bf 61}, 024016 (2000);
B. A. Bassett and F. Tamburini, Phys. Rev. Lett. {\bf 81}, 
2630 (1998).
\bibitem{ABM} 
L. R. Abramo, R. H. Brandenberger, and V. M. Mukhanov, 
Phys. Rev. D {\bf 56}, 3248 (1997).
\bibitem{boson} 
J. Balakrishna, E. Seidel, and W.-M. Suen,  
Phys. Rev. D {\bf 58}, 104004 (1998).
\bibitem{GKLS97}
P. B. Greene, L. Kofman, A. Linde, and A. A. Starobinsky, 
Phys. Rev. D {\bf 56}, 6175 (1997).
\bibitem{carr} 
B. J. Carr and S. W. Hawking, Mon. Not. Roy. Astron. Soc.
{\bf 168}, 399 (1974);
B. J. Carr, Astrophys. J. {\bf 205}, 1 (1975).
\bibitem{BP}
J. S. Bullock and J. Primack, Phys. Rev. D {\bf 55}, 7423 (1997); 
{\it ibid}. astro-ph/9806301.
\bibitem{iv}
P. Ivanov, Phys. Rev. D {\bf 57}, 7145 (1998).
\bibitem{gl} 
A. M. Green and A. R. Liddle, 
Phys. Rev. D {\bf 56}, 6166 (1997);
A. M. Green, A. R. Liddle, and A. Riotto, {\it ibid}.
7554 (1997).
\bibitem{nj} 
J. C. Niemeyer and K. Jedamzik, Phys. Rev. Lett. {\bf 80},
5481 (1998); {\it ibid}. Phys. Rev. D {\bf 59}, 124013 (1999).
\bibitem{latticeeasy} 
G. Felder and L. Kofman, hep-ph/0011160.
\bibitem{eternal}
S. Winitzki and A. Vilenkin, Phys. Rev. D {\bf 61} 084008 (2000).
\bibitem{LL} 
A. R. Liddle and D. H. Lyth, Phys. Rep. {\bf 231}, 1 (1993).
\bibitem{instant}
G. Felder, L. Kofman, A. Linde, Phys. Rev. D{\bf 59}, 123523 (1999).
\bibitem{no}
G. Felder, L. Kofman, A. Linde, Phys. Rev. D{\bf 60}, 103505 (1999).
\end{thebibliography}
\end{document}